# Characterization of the mechanical vibrations of the Schmidt Camera at the Tonantzintla Astronomical Observatory.

**Miguel Antonio Nieto-Suárez**
University of Miami
1320 Campo Sano Drive, Coral Gables, FL, 33124, USA
miguel.nieto@miami.edu

**Daniel Ferrusca R.**
National Institute of Astrophysics, Optics and Electronics (INAOE)
Luis Enrique Erro # 1, Tonantzintla, Puebla, 72840, México
dferrus@inaoep.mx

## ABSTRACT

Mechanical vibrations in astronomical telescopes degrade image quality, making their detection and analysis crucial. This work presents a low-cost, portable vibration monitoring system using MEMS accelerometers and microcontrollers, tested in the historic Schmidt Camera at Tonantzintla. Over 80 years after its commissioning, this is the first study to characterize its mechanical vibrations. Results reveal dominant frequencies and structural couplings that may impact image quality. The system's design, deployment, and preliminary results demonstrate the viability of inertial sensing for heritage-class telescopes and future instrumentation efforts.

For the successful design and integration of new astronomical instruments, a thorough understanding of the host telescope's dynamic environment is critical. Accurate identification and characterization of vibrational patterns allow for the development of corrective techniques and data filtering strategies, ensuring optimal performance. Early knowledge of vibration amplitudes and frequencies also enables predictive modeling of image quality deterioration through numerical simulations, thus informing design decisions from the outset.

This work presents the development and validation of a low-cost, portable vibration detection system based on MEMS (microelectromechanical systems) accelerometers interfaced with a microcontroller. The system architecture including sensor integration, signal acquisition, and data analysis software is described in detail. Laboratory tests were performed to evaluate the system's performance under controlled conditions.

Preliminary on-site results obtained from the Schmidt Camera at the historic Tonantzintla Astronomical Observatory are also presented. Notably, in over 80 years of operation, no prior study had been conducted to characterize the mechanical vibrations affecting the Schmidt Camera. The vibrational analysis offers valuable insights into the telescope's mechanical behavior, identifying dominant frequency components and their corresponding amplitudes. These findings aim to support the preservation of the telescope's operational capabilities and to inform future instrumentation initiatives involving this historic facility.

## 1 INTRODUCTION

The characterization and mitigation of mechanical vibrations in telescopes is a recurring challenge in astronomical instrumentation, particularly in facilities with decades of continuous operation. Such is the case of the Schmidt Camera of Tonantzintla (CST), a historic telescope located in Puebla, Mexico, which has made significant contributions to observational astronomy since its inauguration in 1942. Originally designed to capture wide-field images on photographic plates, the CST has been central to major discoveries, including Herbig–Haro objects, T Tauri stars, and supernovae [1,2]. Today, it continues to operate with digital detectors and a reduced field of view, focusing primarily on the study of minor bodies in the solar system.

Despite its long operational history and various technological upgrades, a detailed study of the mechanical vibrations affecting its structure during tracking or positioning had never been conducted. This work presents a comprehensive evaluation of those vibrations using Inertial Measurement Units (IMUs), devices capable of recording real time acceleration and gyroscopic data. Similar accelerometer-based approaches have been successfully applied to characterize structural behavior in large radio and optical telescopes, such as the Nobeyama 45 m dish [3], the ALMA antennas [4], and the Subaru Telescope [5]. These systems have proven effective in identifying vibration patterns linked to drive motors, wind loads, and mechanical imbalances.

The IMU-based instrumentation used in this study enabled the identification of several vibration sources—such as dome rotation and tracking motors—and the analysis of correlations between vibration intensity and the telescope's pointing direction. The results show that vibrations become more pronounced at low altitudes and to the east near the local meridian, suggesting a structural imbalance. These findings align with previous studies on ground-based telescopes, where resonances in the 10–60 Hz range are often linked to flexure modes or insufficient damping [6,7].

To address these effects, the implementation of mitigation strategies becomes necessary. Contemporary facilities such as the Daniel K. Inouye Solar Telescope (DKIST) and Extremely Large Telescopes (ELTs) have adopted multi-layered vibration control techniques, including feedforward correction using accelerometers, structural stiffening, and modal damping approaches [8,9]. These methods, although more complex, demonstrate the importance of real-time diagnostics and sensor-based feedback in maintaining image quality.

In this context, IMUs offer a cost-effective and adaptable solution, particularly valuable for historical telescopes with limited infrastructure. Previous publications have highlighted the potential of these sensors for observatory applications [10]. Furthermore, the present study builds on earlier work by the authors focused on the integration of inertial systems in astronomical telescopes [11], as well as resources made available through the ScopeVibes project, an ongoing initiative dedicated to structural monitoring in observatories [12].

Although image degradation in ground-based astronomy is often attributed solely to atmospheric turbulence ("seeing"), it is crucial to recognize other contributing factors, such as mechanical vibrations of the telescope structure itself. These vibrations can introduce additional blur or distortion, especially when real-time correction systems are not available. In such cases, acknowledging and quantifying mechanical disturbances becomes essential to filter out the most degraded frames prior to image stacking and processing, thereby improving the final data quality.

This study represents an initial step toward the modernization of structural diagnostics in legacy telescopes, emphasizing the practicality and effectiveness of inertial sensors as tools for vibration assessment in observational astronomy.

## 2 MATERIALS AND METHODS

The application of inertial sensors for structural diagnostics has proven highly effective across multiple engineering domains. In particular, Inertial Measurement Units (IMUs)—which integrate triaxial accelerometers and gyroscopes—enable real-time acquisition of linear acceleration and angular velocity. When applied to astronomical telescopes, IMUs allow the detection and quantification of vibrational motion across multiple axes, providing insight into mechanical behavior under operational conditions.

For this study, we selected the MPU6050 [13], a compact MEMS-based sensor combining a 3-axis accelerometer and 3-axis gyroscope. The sensor was interfaced with an Arduino Uno R3 microcontroller [14] via the I²C protocol, which also permits the integration of multiple sensors. Figure 1 illustrates the wiring configuration for dual MPU6050 modules. This dual-sensor setup provides redundancy, reducing random noise and enhancing the signal-to-noise ratio. Additional implementation details can be found in [12].

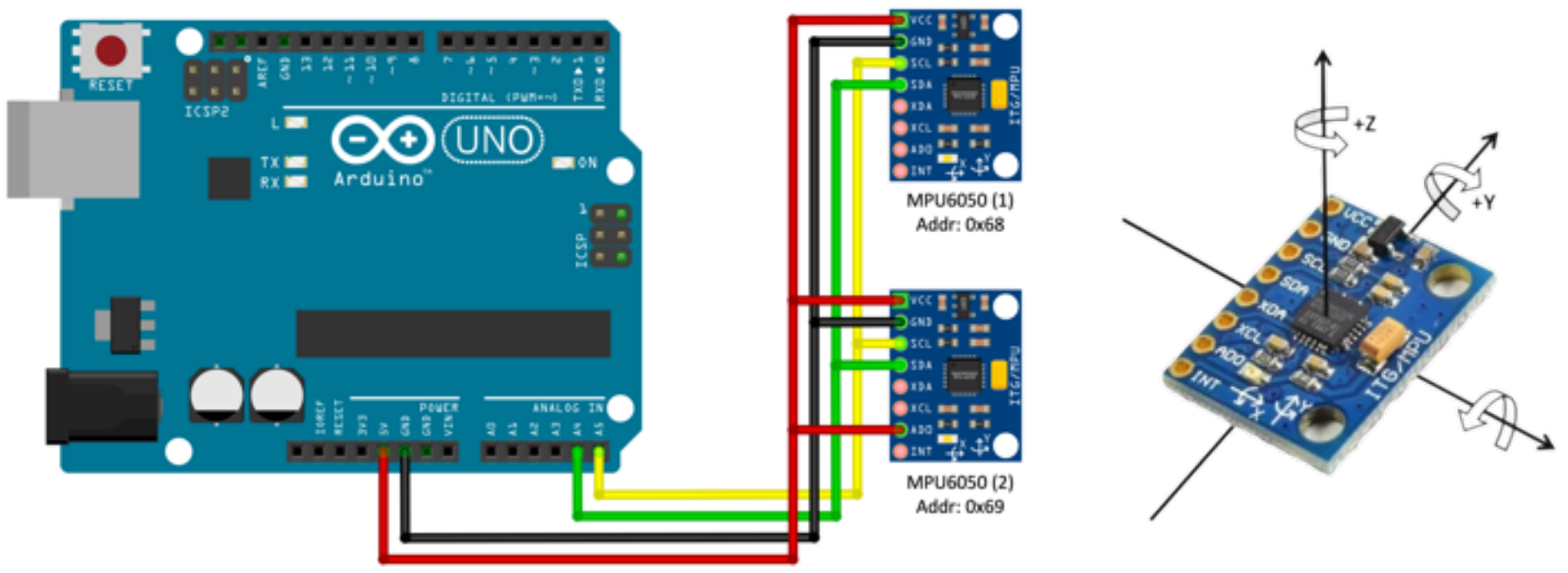

**Figure 1**. Left panel: Connection diagram of two MPU6050 sensors to an Arduino Uno via I²C. Different addresses (0x68, 0x69) are set through the AD0 pin, enabling redundant data acquisition to improve signal quality, created with Fritzing. Right panel: Representation of the sensor axes.

To facilitate secure and modular deployment on the telescope structure, the IMUs and microcontroller were enclosed in custom 3D-printed housings, incorporating RJ11 connectors for reliable 4-wire communication and neodymium magnets to ensure rigid attachment to ferrous surfaces (as shown in Figure 2). The Arduino Uno and sensors were powered via the USB data cable itself, whose length should not exceed 10 meters to ensure compliance with the required data transfer rate specifications.

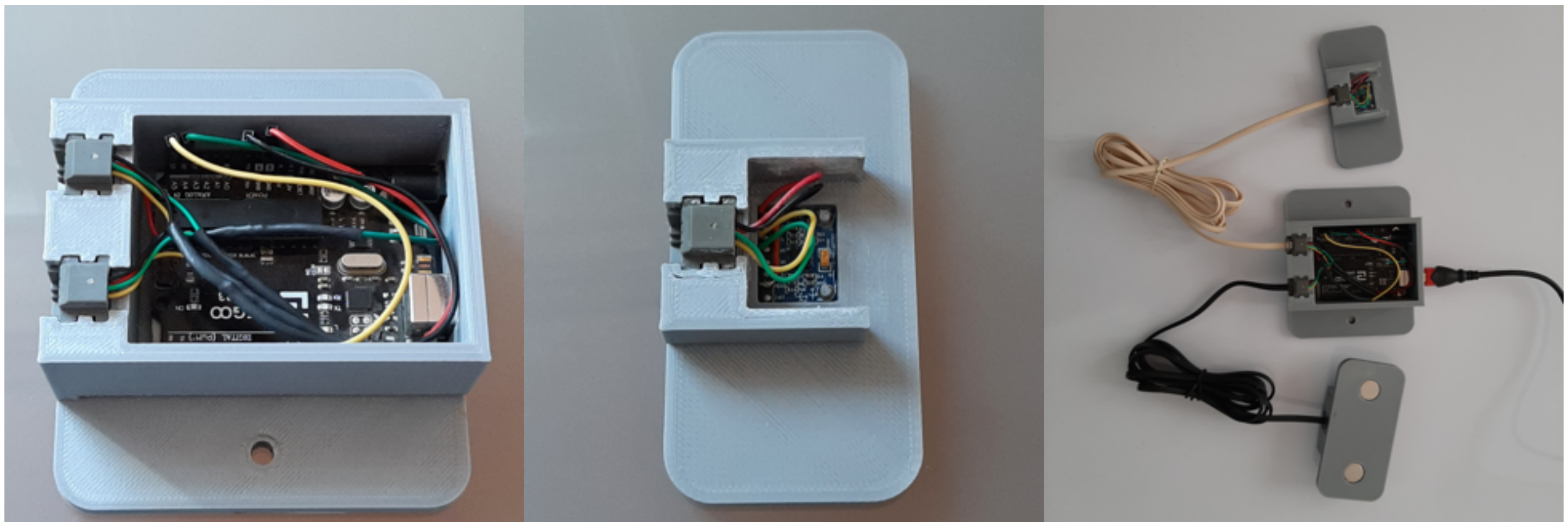

**Figure 2**. Arduino Uno (left panel) and MPU6050 (center panel) sensors housed in custom 3D-printed enclosures, equipped with neodymium magnets to ensure robust mechanical coupling. Both modules are interconnected via RJ11 connectors (right panel).

This communication between the Arduino and a control PC was established at 115200 baud, with data acquisition managed by a custom MATLAB script. The sampling rate was experimentally varied between 30 Hz and 1 kHz and ultimately fixed at 350 Hz for spectral analysis, after in situ verification confirmed that the telescope was not affected by high-frequency vibrations. This sampling rate provided optimal resolution for the frequency range of interest.

Each data acquisition session was stored in files containing matrices of 16,384 rows by 6 columns. The 16,384 rows ($2^{14}$), a power-of-two length selected to optimize Fast Fourier Transform (FFT) efficiency in the subsequent spectral analysis. Each row comprised six channels of inertial measurements obtained from the MPU6050 sensor, which integrates a triaxial accelerometer and gyroscope. The variables **AccX**,

**AccY**, and **AccZ** correspond to the linear acceleration components along the orthogonal axes X, Y, and Z, respectively, while **GyrX**, **GyrY**, and **GyrZ** represent the angular velocity components about the same axes. These six-degree-of-freedom signals enable precise characterization of both translational and rotational dynamics, which are critical for vibration diagnostics and structural response monitoring. The total number of files then depends on the data acquisition time, which can be subsequently processed individually or averaged with other files, drastically reducing the level of random noise. By calculating the FFT of each column of this matrix, the frequency spectrum of the vibration component on each axis is obtained.

If the time series data stored in each column, such as the three corresponding to the acceleration components along the X, Y, and Z axes measured by the MPU6050, are plotted against time and subsequently subjected to a Fourier Transform, the resulting spectra reveal the amplitude of acceleration variations as a function of frequency for each axis. This procedure is illustrated in Figure 3 using MATLAB. In the left panel, time-domain acceleration data from an IMU positioned horizontally on a speaker diaphragm are shown. The speaker, also oriented horizontally, was driven by a signal generator producing a 20 Hz tone. The abscissa axis, initially labeled as "Number of data," can be readily converted to time in seconds based on the known sampling rate of 350 Hz. The right panel displays the corresponding frequency spectra for all three axes. Notably, the Z-axis exhibits the highest amplitude, consistent with the direction of vibration of the speaker cone. This and similar experiments served to validate the frequency calibration of both the instrumentation and the signal processing code.

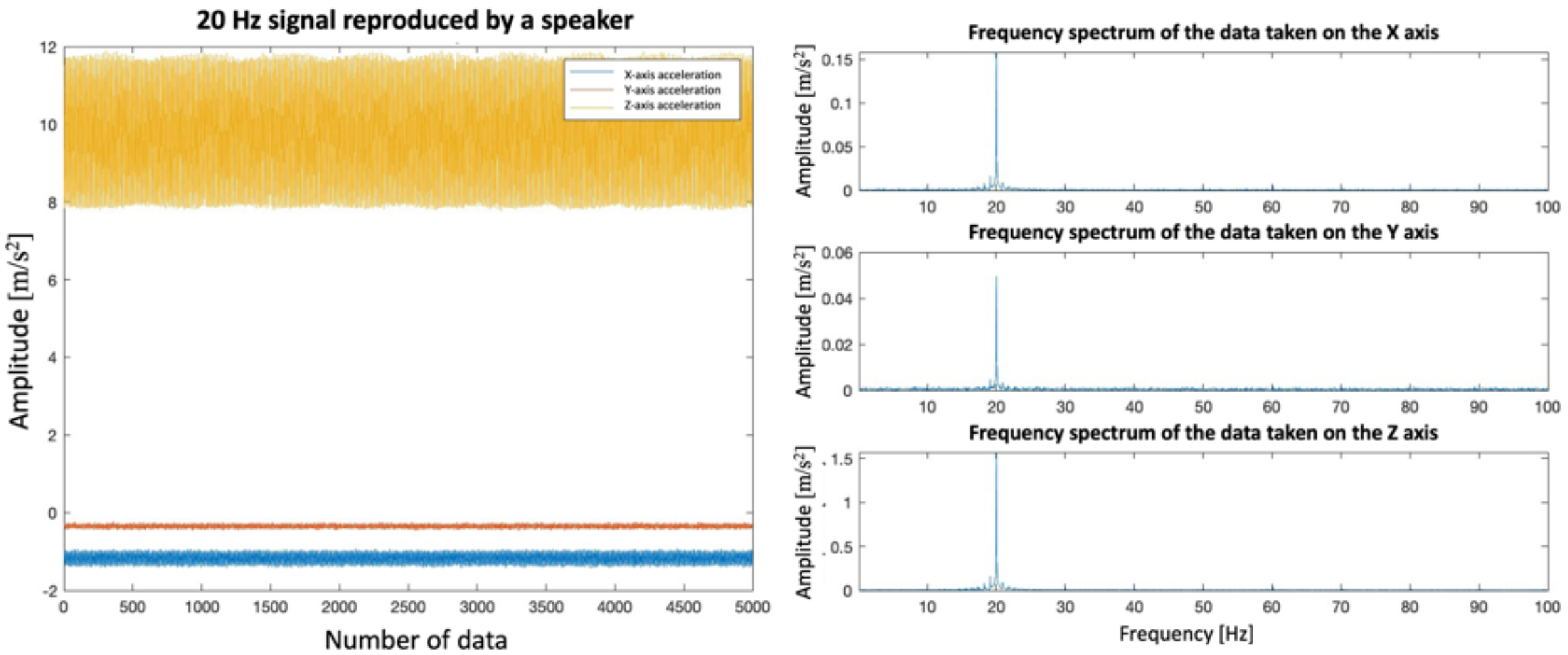

**Figure 3**. Time- and frequency-domain analysis of acceleration signals recorded with an MPU6050 sensor placed on a speaker diaphragm emitting a 20 Hz tone. Left panel: Raw acceleration data. Right panel: Frequency spectra showing a dominant 20 Hz peak, with the Z-axis exhibiting the highest amplitude due to alignment with the vibration direction.

The MPU6050 sensors were strategically installed at different locations on the telescope, as shown in Figure 4. (1) Two sensors were placed at the base of the telescope, where the greater distance from the support fork and the higher moment of inertia allow for better detection of high-energy vibrations. (2) A sensor was positioned inside the telescope, adjacent to the detector, to capture vibrations that most significantly degrade astronomical images. Additional sensors were mounted (3) near the declination motor and (4) the right ascension motor to monitor vibrations generated by these components. Furthermore, sensors were also placed on the dome traction motors and on the air conditioning compressor located at the base of the CCD camera.

In parallel, to confirm whether the vibrations detected in the telescope could have affected astronomical images already taken, hundreds of raw images acquired with the CST using various filters, on different nights and at multiple times, were analyzed. These were compared with stacked and processed images using the IRAF (Image Reduction and Analysis Facility) software package as well as custom MATLAB scripts developed in-house.

Finally, a detailed inspection of the observatory structure and the telescope's mechanical components was carried out to identify potential sources of vibration and their propagation mechanisms. External factors such as wind speed were not considered during observation nights, as standard protocols for dome closure under adverse weather conditions were strictly followed.

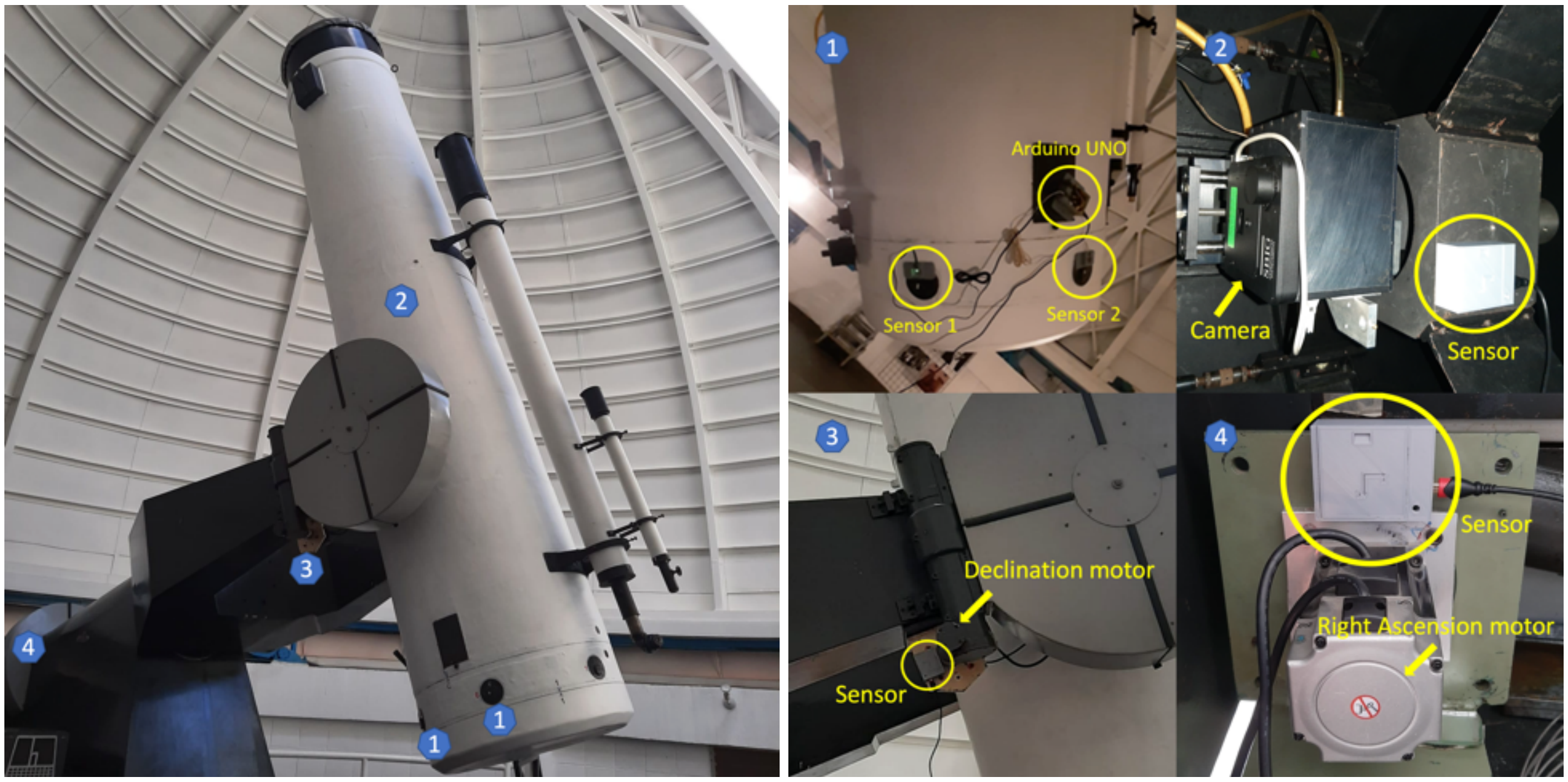

**Figure 4**. Left panel: Locations on the telescope where MPU6050 sensors were strategically placed: (1) at the telescope base, (2) inside the telescope near the detector, (3) adjacent to the declination motor, and (4) adjacent to the right ascension motor. Right panel: Close-up views of the four selected sensor positions.

**Note**: All instrumentation setups, firmware, and data analysis code developed for this study are open-source and may be provided to interested researchers upon reasonable request.

## 3 RESULTS

This section presents a multi-faceted analysis of the mechanical vibrations affecting the Schmidt Camera at the Tonantzintla Astronomical Observatory. First, in situ measurements using MPU6050 sensors placed at strategic locations on the telescope structure reveal the presence and propagation of vibrations induced by dome motion and internal mechanisms, with notable dependence on the telescope's pointing elevation. Second, the analysis of raw astronomical images uncovers systematic elongation in stellar profiles, suggesting mechanical contributions to image degradation. Finally, a structural assessment identifies unintended mechanical coupling and suboptimal motor placement as key factors contributing to vibrational transmission, offering specific targets for mitigation to improve observational stability.

### 3.1 Analysis of Vibration Signatures Across Telescope Structure

Two MPU6050 sensors were mounted at the base of the telescope (position 1 in Figure 4), and data were recorded over an 8-minute period while both the telescope and the dome remained stationary. With the telescope pointed at the zenith, this configuration served as a reference to characterize the baseline noise level (left panel of Figure 5). Subsequently, additional measurements were acquired over a 2-minute interval during continuous dome rotation, operating at approximately 1 RPM. A clear increase in vibration levels induced by dome motion is evident under these conditions (right panel of Figure 5). This procedure was repeated at various telescope elevation angles (40°, 50°, 60°, 70°, 80°, and 90°), all oriented toward the south (azimuth 180°). The results, presented in Figure 6, indicate that vibrations induced by dome rotation are more pronounced at lower elevation angles.

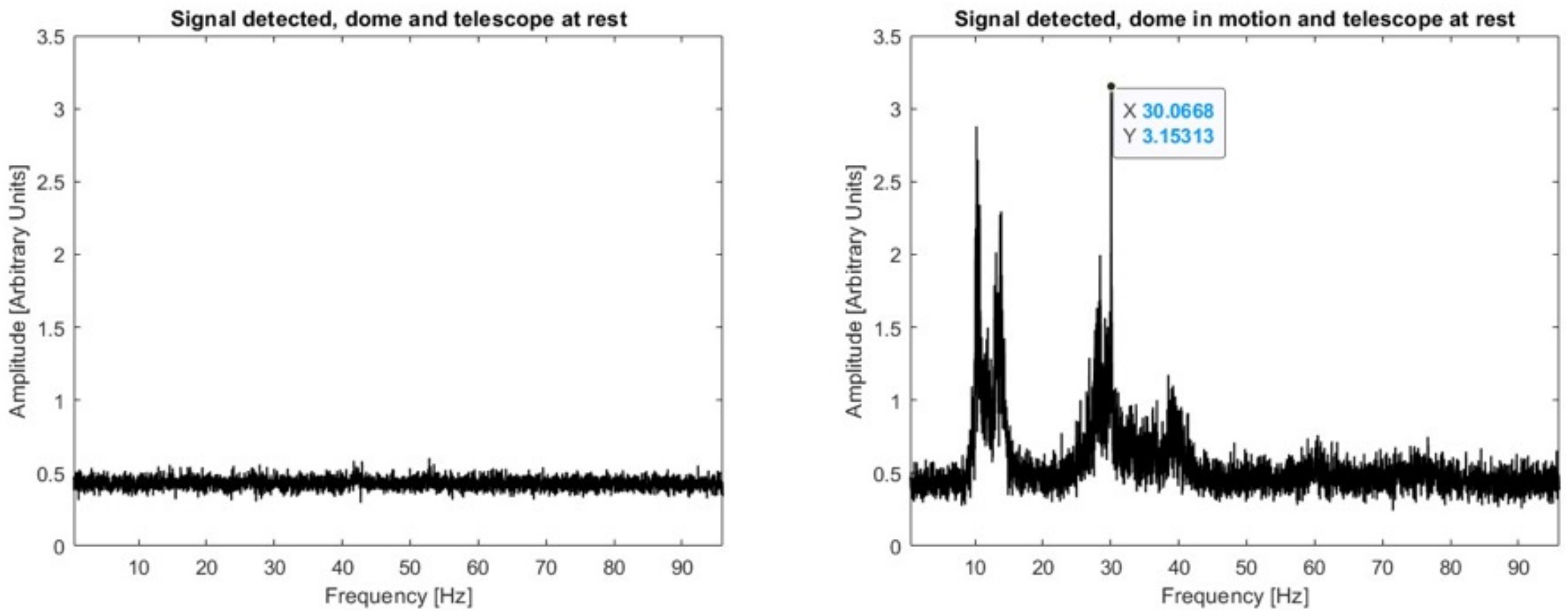

**Figure 5**. Left panel: no significant vibrations detected with both the telescope and dome at rest. Right panel: vibrations appear at the telescope base during dome motion.

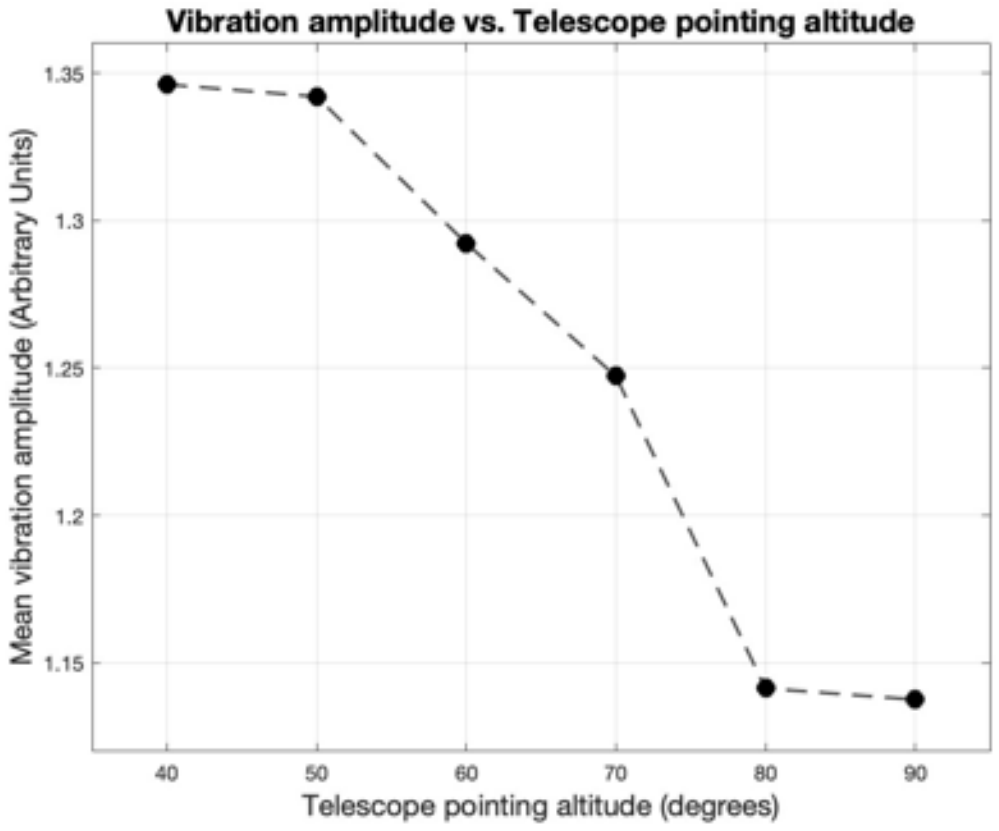

**Figure 6**. Average vibration amplitude at the telescope during dome rotation for different elevation angles. Vibrations are more pronounced at lower elevations, indicating a stronger mechanical coupling between dome motion and the telescope structure at these positions.

The analysis of vibrations generated by the three motors responsible for dome rotation, focusing in particular on the spectral profile of motor 1 (Figure 7), which is located closest to the telescope base (see Figure 14), revealed a clear similarity to the vibration spectrum recorded at the telescope during dome motion (right panel of Figure 5). The dominant spectral peak at 30 Hz corresponds to the motor's nominal rotational speed of 1800 RPM, as indicated on the manufacturer's nameplate. Additional frequency components observed at the motor are attributed to mechanical resonances governed by the geometry and material properties of the support frame. A comparison between Figures 7 and 5 (right panel), shows that higher-frequency components are substantially attenuated by the dome structure, resulting in minimal transmission of those vibrations to the telescope. While the vibration amplitude at the telescope base is substantially attenuated compared to that measured at the motor, with a reduction factor greater than 20, it remains within a range that could adversely affect image quality, particularly when the telescope is positioned at low elevation angles where structural sensitivity to vibrations is increased.

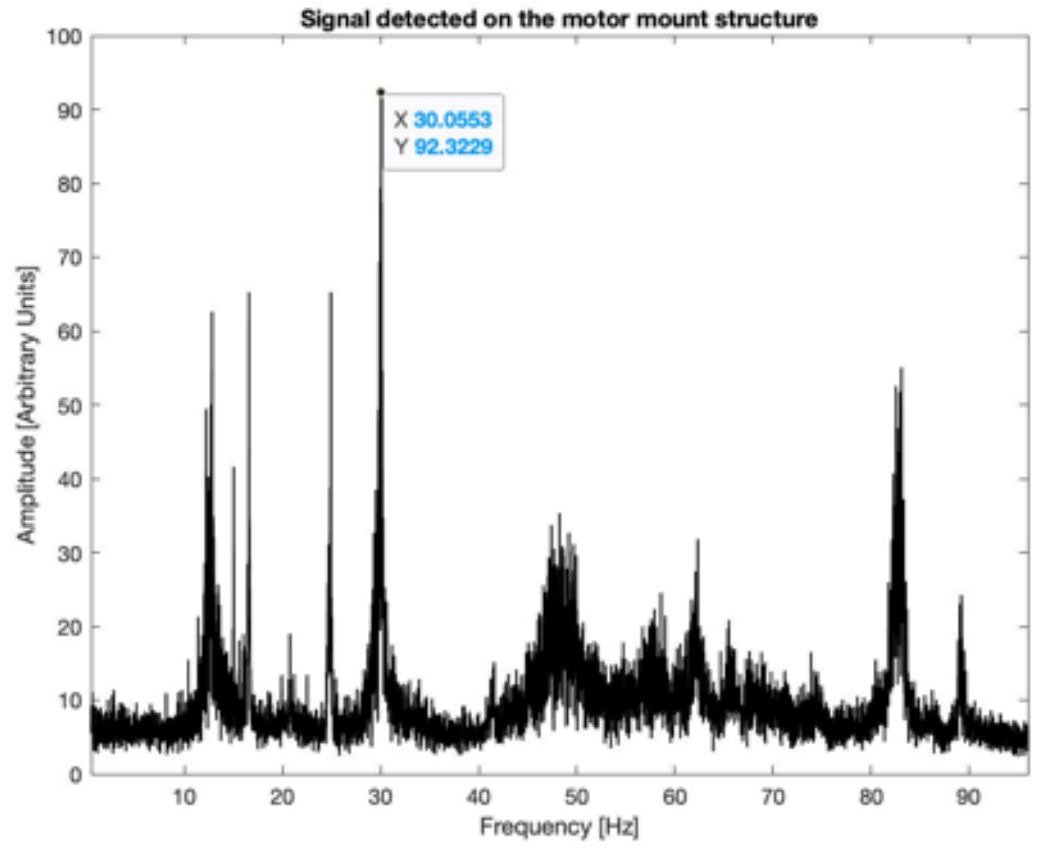

**Figure 7**. Vibration spectrum measured directly on motor 1 of the dome rotation system. The dominant peak at 30 Hz corresponds to the motor's nominal speed of 1800 RPM. Additional peaks are associated with structural resonances of the motor mount.

To investigate vibrations affecting the optical detector directly, an MPU6050 sensor was attached adjacent to the camera (position 2 in Figure 4). With the dome closed and stationary, the telescope was programmed to simulate sidereal tracking by sweeping across hour angles from –3h to +3h using the right ascension (RA) motor, for different declination values. For example, at a declination of 19° N, the telescope points toward the zenith (altitude of 90° above the horizon) when crossing the local meridian (hour angle 0h), which corresponds to the geographical latitude of the Schmidt Camera site (19° N). For a declination of 26° S, the telescope reaches an altitude of 45° above the horizon when aligned with the meridian. Since the telescope is mounted equatorially, once a celestial object is acquired using both the RA and declination motors, only the RA motor remains active during continuous tracking.

It is noteworthy that during the telescope's sidereal tracking motion, both at declinations of 19°N and 26°S, the maximum amplitude of the detected 4.5 Hz vibrations occurs near 0:00 hours hour angle, that is, when the telescope is pointing directly south. Notably, two vibration peaks are observed: one at 3.75 degrees east and another at 7.5 degrees west of the local meridian. This slight asymmetry in the vibration pattern suggests the presence of a minor mechanical imbalance in the telescope structure (see Figure 9).

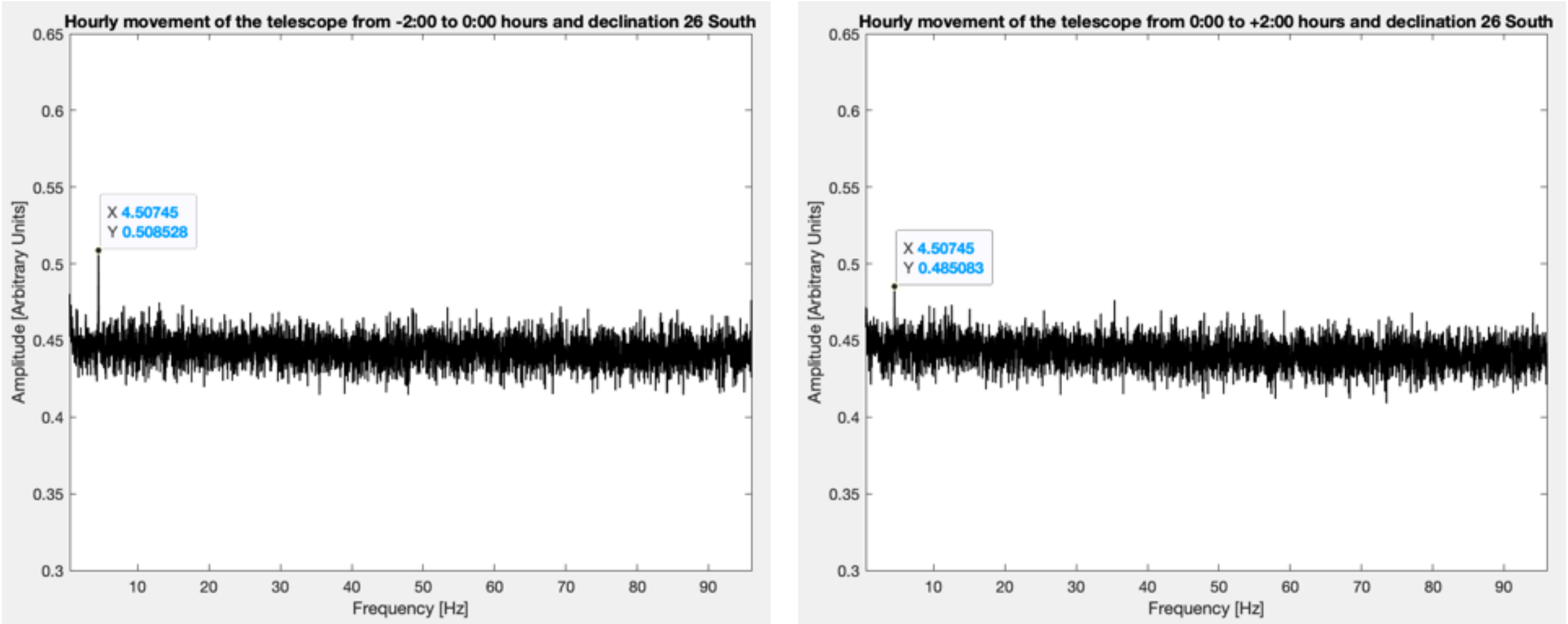

**Figure 8**. Average frequency spectra of telescope motion at a declination of 26° S. Left panel: spectra from –2h to 0h; right panel: spectra from 0h to +2h. A prominent 4.5 Hz vibration is present in both cases, with higher amplitudes than those observed at 19° N.

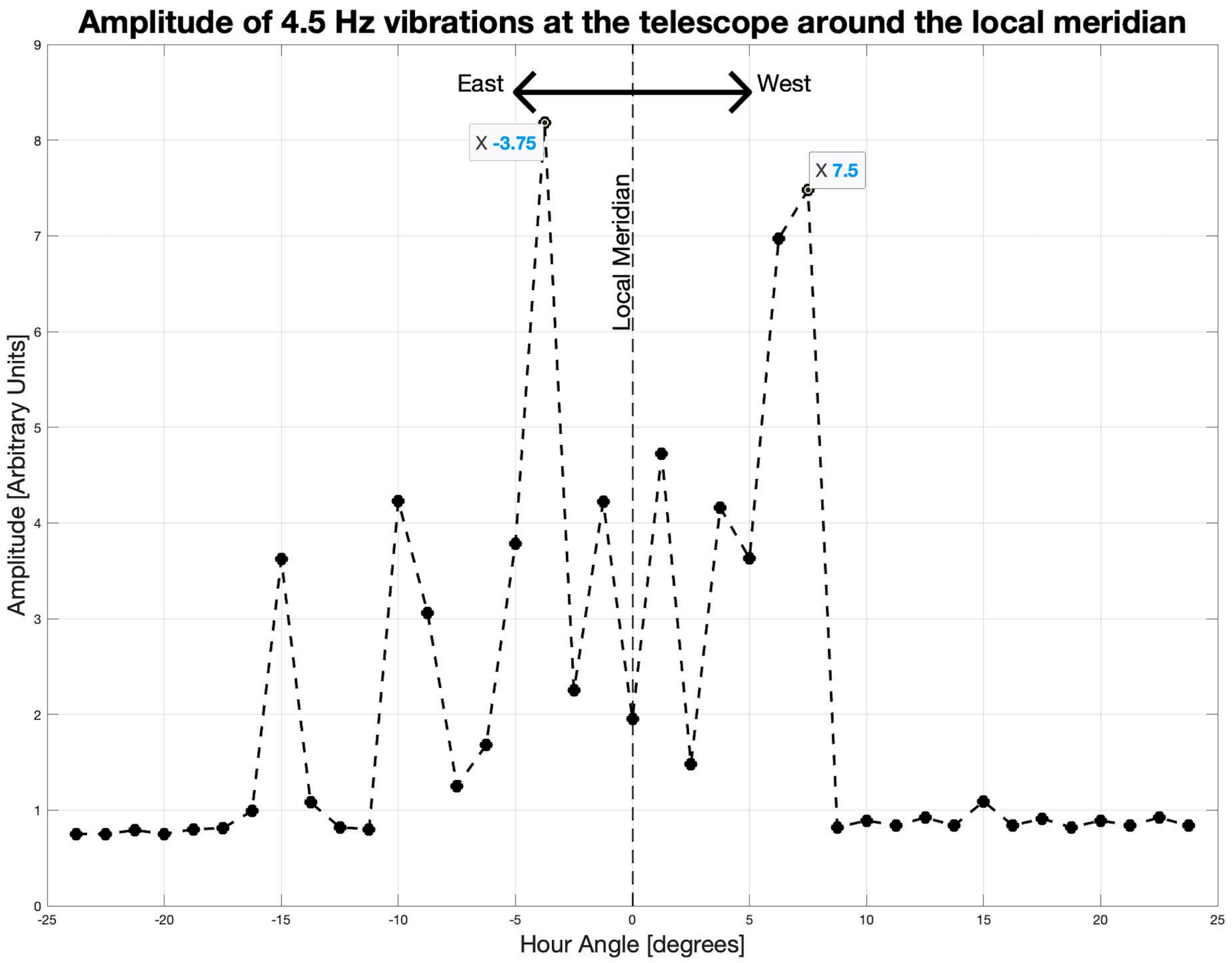

**Figure 9**. Variation in the amplitude of the 4.5 Hz vibrations detected during the telescope's sidereal tracking, from −23.75° (1 h 35 min east) to +23.75° (1 h 35 min west). The results show peak amplitudes on both sides of the meridian pointing, although not symmetrically.

## 3.2 Analysis of Stellar Profiles in Raw Images Acquired with the Schmidt Camera

It is crucial to understand the potential consequences of the detected vibrations described in section 3.1, as these vibrations can directly impact the optical performance and image quality of the telescope. This section presents an analysis of the stellar profiles observed in raw images obtained with the Schmidt Camera, aiming to correlate the mechanical disturbances with tangible effects on the astronomical data.

An inspection of hundreds of raw astronomical images previously taken at different epochs through different filters using the Schmidt Camera reveals a consistent asymmetry in the stellar profiles across the field of view, with the horizontal axis systematically broader than the vertical one (see Figure 10). This effect becomes more evident when zooming in on individual stars (see Figure 11), where it is clear that the point sources lack a circularly symmetric profile, as would be expected under purely atmospheric seeing conditions. In such cases, the Point Spread Function (PSF) should exhibit an isotropic broadening, characterized by a Full Width at Half Maximum (FWHM) that is statistically similar in all directions. The observed elongation along a preferred axis instead suggests the influence of mechanical tracking errors or structural vibrations of the telescope, likely affecting that specific direction.

These findings point to a clear link between the mechanical behavior of the telescope and the quality of the recorded images. However, the elongation observed in the stellar profiles from the raw image analysis appears to result from some form of degradation, either due to imperfections in the sidereal tracking motion or to the influence of mechanical vibrations. Notably, when all astronomical images of the same field of view are stacked and reduced, this elongation is significantly diminished. This behavior suggests that image defects occurring before the telescope crosses the local meridian may be partially compensated for by those acquired after crossing it. To validate this hypothesis, a more precise and systematic investigation is required, underscoring the need for continued studies on the temporal and directional dependence of the detected vibrations and their effects on image quality.

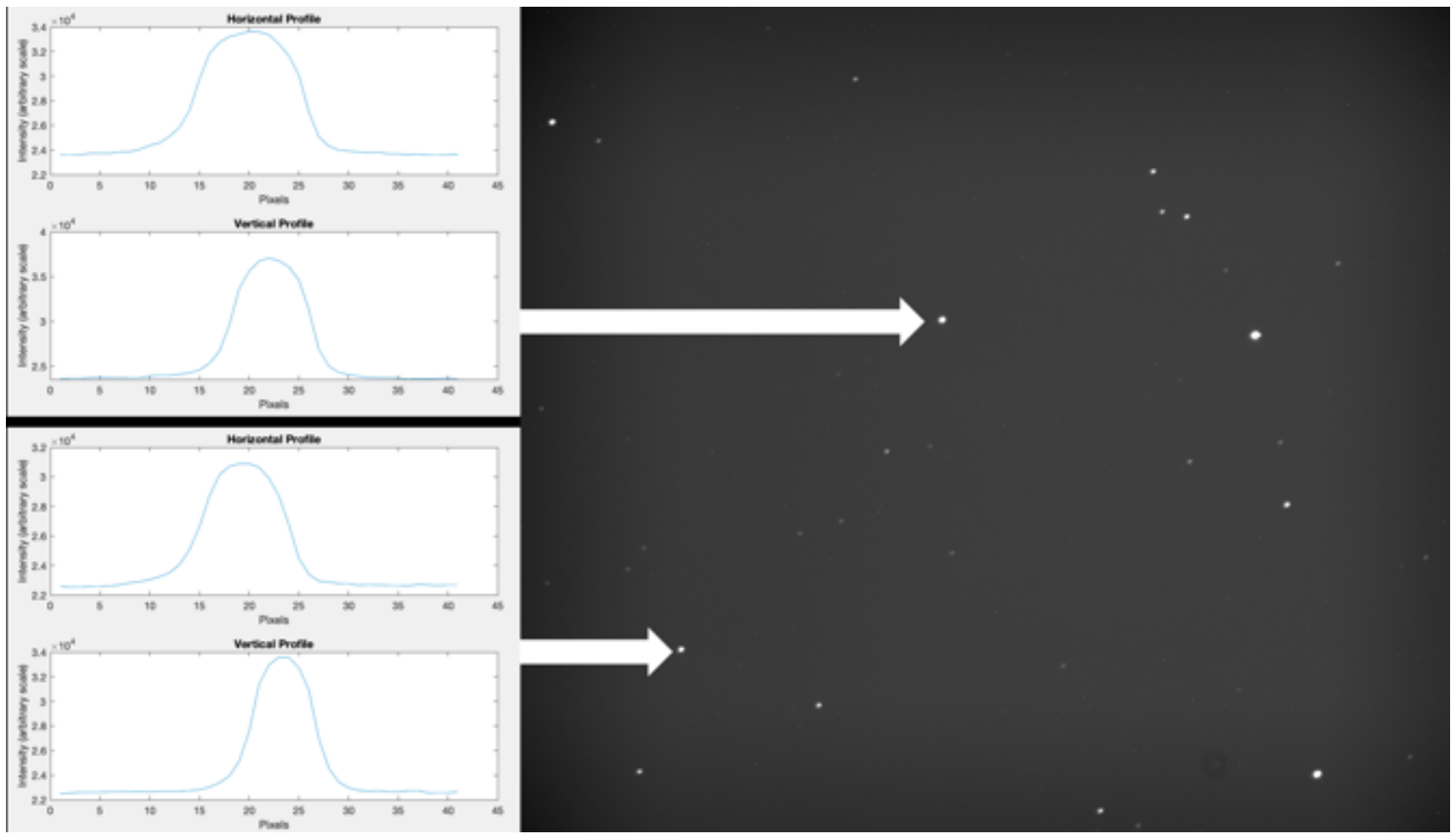

**Figure 10**. Horizontal and vertical profiles of stars in raw Schmidt Camera images. A consistent broadening is observed along the horizontal axis.

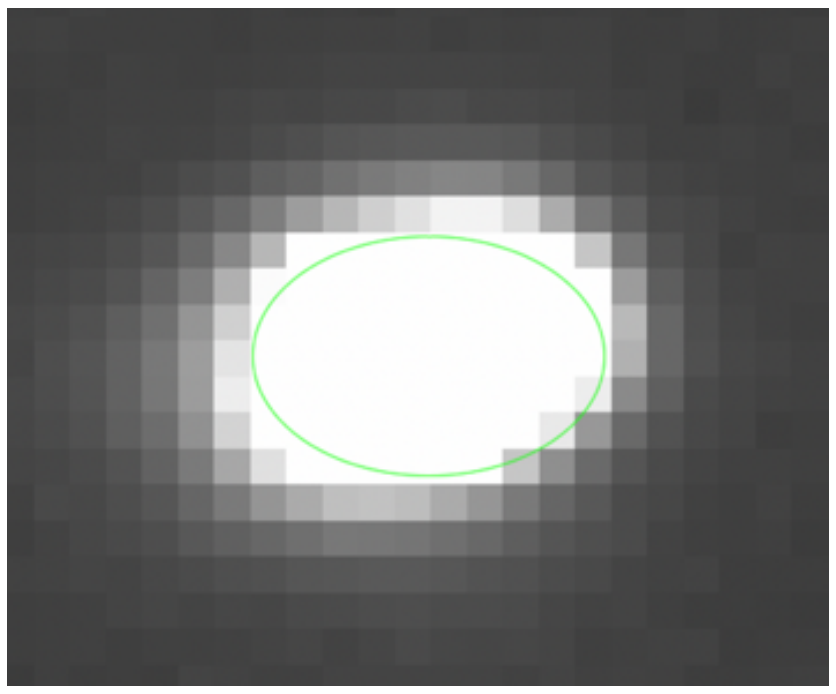

**Figure 11**. Zoom-in of a representative star. The elongation suggests tracking errors or telescope vibrations. The small squares in the image correspond to pixels.

### 3.3 Analysis of Structural Coupling and Motor Placement Effects

Given the importance of understanding and mitigating mechanical vibrations to preserve image quality, this section emphasizes the critical need to verify the possible consequences of the vibrational sources identified earlier (section 3.1). Detailed structural inspection and evaluation of motor placement provide insights into unintended vibration pathways and mechanical weaknesses.

An unexpected mechanical coupling was identified between the telescope base and the dome floor. At some point in the history of the Schmidt Camera, maintenance personnel appear to have sealed the gap between the floor and the base of the telescope, possibly for aesthetic reasons, without awareness of its critical role in isolating the telescope from structural vibrations originating in the building (Figure 12). This unintentional contact is compromising the mechanical decoupling designed to preserve the precision and stability of astronomical observations. The finding underscores the importance of detailed inspections of the telescope's mechanical interfaces and support structures, as such oversights could significantly impact observational performance.

Additionally, electronic instrumentation was found mounted on the lower floor of the observatory and directly attached to the telescope base (Figure 12, right panel). Although the vibrational energy introduced by this equipment may be negligible relative to the telescope's total mass, it nonetheless constitutes a preventable source of mechanical noise. Relocating this instrumentation to another wall of the building is expected to reduce potential coupling and improve the telescope's mechanical stability during operation.

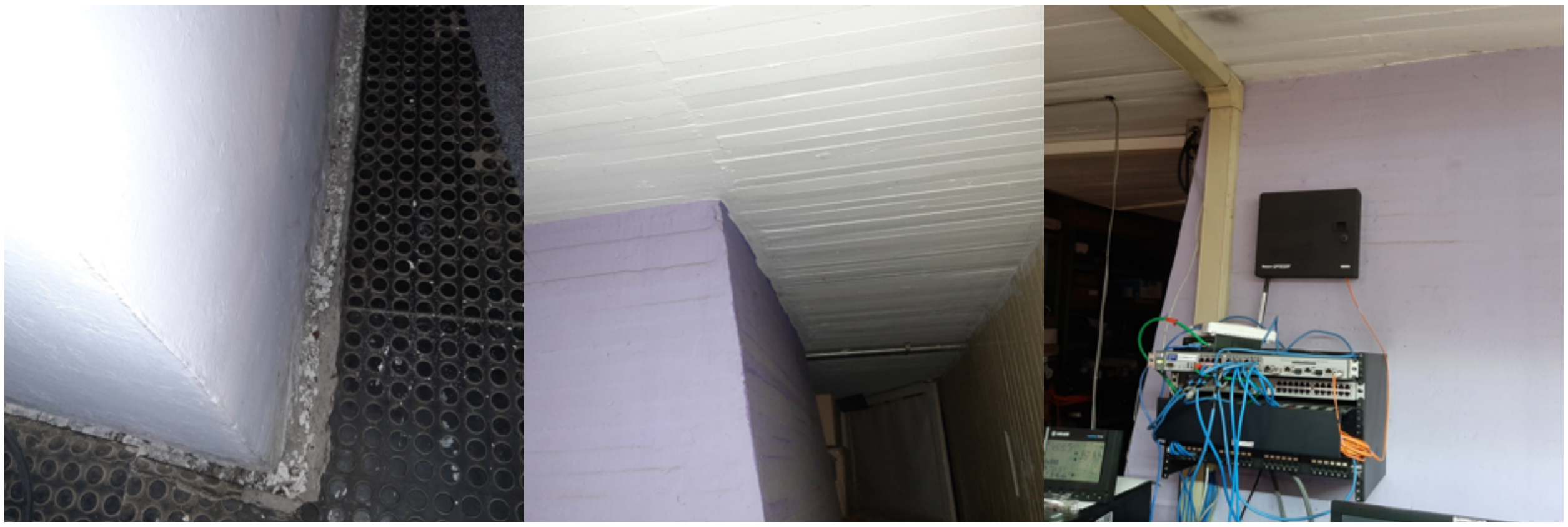

**Figure 12**. Left panel: View of the telescope base and dome floor from above, showing direct contact. Center panel: View from the lower floor, highlighting the filled gap between the telescope base and the building structure. Right panel: Electronic instrumentation mounted directly on the telescope base, identified as a secondary source of vibrations.

The telescope dome is supported by ten bearings, three of which are coupled to electric motors responsible for driving its rotation (see schematic in Figure 13). Both the free-rolling and motorized bearings are distributed as evenly as possible around the perimeter to ensure balanced load support and uniform traction. However, the drive bearing associated with motor 1 is located very close to the telescope base and is likely the primary source of the vibrations transmitted to the structure. Repositioning each motor to the next bearing in the counterclockwise direction would increase the separation between motor 1 and the telescope base, placing all three motors at more comparable distances from the telescope. Additionally, this rearrangement would allow the drive system to operate through less worn bearings, potentially reducing vibration transmission and improving mechanical performance.

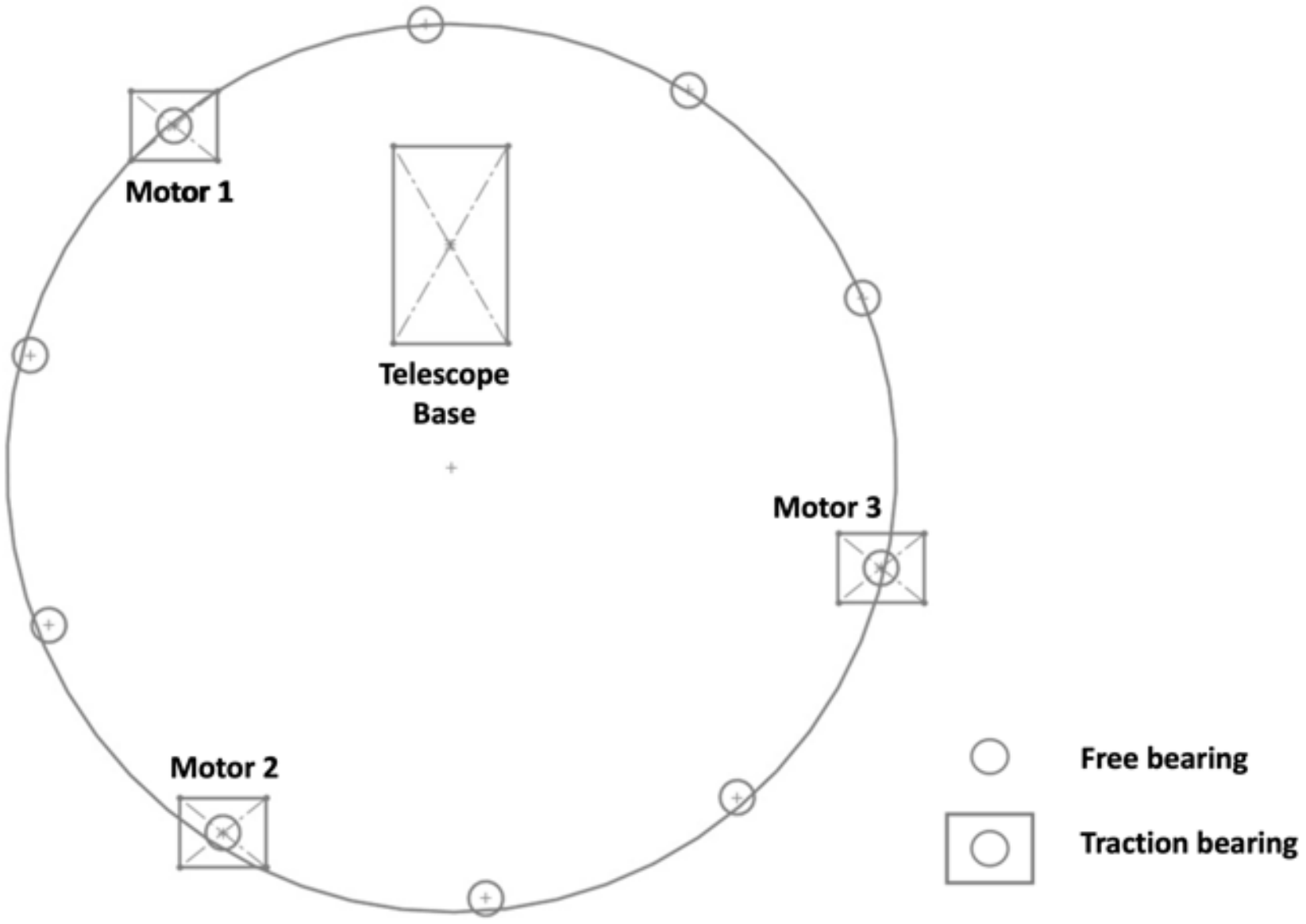

**Figure 13**. Schematic of the dome and the distribution of its support bearings. The first free-rolling bearing, located directly in front of the telescope base, is misaligned with respect to it. Additionally, the drive bearing connected to motor 1 is positioned extremely close to the telescope base.

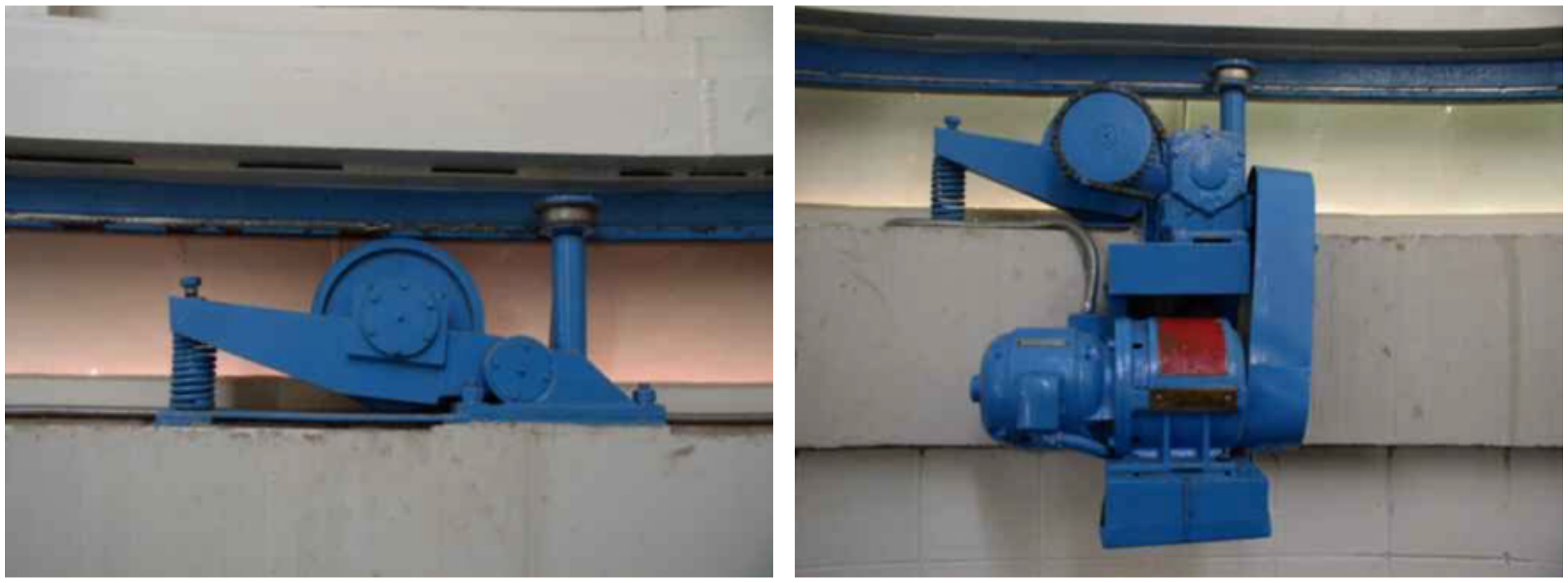

**Figure 14**. Left panel: free-rolling bearing. Right panel: motorized drive bearing.

## 4 CONCLUSIONS

This study represents an important first step in the systematic analysis of mechanical vibrations affecting the Schmidt Camera Telescope of Tonantzintla, a historically and scientifically valuable astronomical instrument. By employing Inertial Measurement Units (IMUs), we successfully recorded and characterized dynamic disturbances compromising the telescope’s structural stability and, consequently, the quality of its astronomical observations.

The data revealed several sources of vibration, most notably those associated with the dome rotation motors and the telescope tracking system. These vibrations are not uniformly distributed; they intensify at lower pointing altitudes and in specific orientations relative to the local meridian. This asymmetric behavior points to a slight mechanical imbalance in the telescope structure and highlights a direct relationship between the telescope's mechanical configuration and the degradation of its optical performance.

The use of IMUs proved to be a key component in the dynamic assessment of the system. Their ability to capture acceleration across multiple axes provided not only the detection of vibrations but also insight into their temporal and spatial evolution. Such information is essential for informed operational decisions, including the validation or rejection of astronomical images affected by excessive vibrational noise.

Based on our findings, we propose targeted actions to mitigate the effects of these vibrations: decoupling the telescope base from the dome platform to reduce mechanical transmission; rotating the dome traction motors to less worn bearing zones to reduce instability and distance the most disruptive motor from the telescope base; and correcting the detected structural imbalance. We also recommend the permanent integration of IMU-based monitoring systems to provide real-time feedback to operators regarding vibrational conditions during observations.

In addition, the device developed in this study offers significant advantages compared to commercial solutions. It is a low-cost, open-source instrument combined with custom analysis software, demonstrating full functionality for vibration monitoring in astronomical telescopes. The open-source nature facilitates easy replication and customization, enabling the potential creation of a distributed network of sensors for comprehensive monitoring of telescopes or other sensitive environments. This capability makes the system not only a practical tool for the Schmidt Camera but a promising template for similar applications in heritage-class and modern observatories worldwide.

Taken together, this work demonstrates the potential of inertial sensing as a practical, cost-effective, and robust approach for vibrational diagnostics in heritage-class telescopes. Its application preserves the quality of observational data while extending the functional lifetime of the instrument, ensuring its continued contribution to astronomical research.

# 5 ACKNOWLEDGMENTS

We acknowledge the collaboration grant from INAOE awarded to Dr. Miguel Antonio Nieto-Suárez, as well as the support provided by the technicians of the Schmidt Camera in Tonantzintla, Agustín Márquez Limón and Guillermo Cerdán Hernández, which was essential for the realization of this study.